\documentclass[twocolumn]{aastex631}
\usepackage{amsmath}

\received{2023 July 13}
\revised{2023 September 23}
\accepted{2023 September 25}

\submitjournal{ApJL}

\begin{document}

\title{Turbulence Properties of Interplanetary Coronal Mass Ejections in the Inner Heliosphere: Dependence on Proton Beta and Flux Rope Structure}

\correspondingauthor{S. W. Good}
\email{simon.good@helsinki.fi}

\author[0000-0002-4921-4208]{S. W. Good}
\affil{Department of Physics, University of Helsinki, PO Box 64, FI-00014 Helsinki, Finland}

\author[0000-0001-7301-2617]{O. K. Rantala}
\affil{Department of Physics, University of Helsinki, PO Box 64, FI-00014 Helsinki, Finland}

\author[0000-0002-2918-5120]{A.-S. M. Jylh\"{a}}
\affil{Department of Physics, University of Helsinki, PO Box 64, FI-00014 Helsinki, Finland}

\author[0000-0003-4529-3620]{C. H. K. Chen}
\affil{Department of Physics and Astronomy, Queen Mary University of London, London E1 4NS, UK}

\author[0000-0001-6868-4152]{C. M\"ostl}
\affil{Austrian Space Weather Office, GeoSphere Austria, Reininghausstrasse 3, 8020 Graz, Austria}

\author[0000-0002-4489-8073]{E. K. J. Kilpua}
\affil{Department of Physics, University of Helsinki, PO Box 64, FI-00014 Helsinki, Finland}

\begin{abstract}

Interplanetary coronal mass ejections (ICMEs) have low proton beta across a broad range of heliocentric distances and a magnetic flux rope structure at large scales, making them a unique environment for studying solar wind fluctuations. Power spectra of magnetic field fluctuations in 28 ICMEs observed between 0.25 and 0.95~au by Solar Orbiter and Parker Solar Probe have been examined. At large scales, the spectra were dominated by power contained in the flux ropes. Subtraction of the background flux rope fields reduced the mean spectral index from $-5/3$ to $-3/2$ at $kd_i \leq 10^{-3}$. Rope subtraction also revealed shorter correlation lengths in the magnetic field. The spectral index was typically near $-5/3$ in the inertial range at all radial distances regardless of rope subtraction, and steepened to values consistently below $-3$ with transition to kinetic scales. The high-frequency break point terminating the inertial range evolved approximately linearly with radial distance and was closer in scale to the proton inertial length than the proton gyroscale, as expected for plasma at low proton beta. Magnetic compressibility at inertial scales did not show any significant correlation with radial distance, in contrast to the solar wind generally. In ICMEs, the distinctive spectral properties at injection scales appear mostly determined by the global flux rope structure while transition-kinetic properties are more influenced by the low proton beta; the intervening inertial range appears independent of both ICME features, indicative of a system-independent scaling of the turbulence.

\end{abstract}


\section{Introduction} \label{sec:intro}

Alongside the continuous outflows of the fast and slow solar winds, interplanetary coronal mass ejections \citep[ICMEs;][]{Kilpua17} represent a third, impulsive type of solar wind in the heliosphere. The plasma, magnetic field, and compositional properties of ICMEs differ to those of the other solar wind types in various significant ways \citep{Zurbuchen06}, making ICMEs a distinctive environment for investigating a range of space plasma phenomena.

A characteristic signature of ICME wind at 1~au is the dominance of magnetic pressure over the ion component of the plasma pressure (i.e. $\beta_i \ll 1$), a result of strong magnetic fields combined with low proton temperatures. Strong fields in ICMEs are often associated with the presence of large-scale, nearly force-free magnetic flux ropes \citep{Goldstein83}. An ICME is the interplanetary manifestation of an erupted coronal flux rope, although ICMEs are not always observed to have a flux rope structure in situ \citep[e.g.][]{Cane03}. ICMEs typically have field strengths in excess of the ambient solar wind field across a wide range of heliocentric distances \citep[e.g.][]{Wang05}. Low proton temperatures and the absence of a proton temperature-velocity correlation in ICMEs have been attributed to rapid expansion close to the Sun \citep[e.g.][]{Matthaeus06} and to the dominance of magnetic forces in the magnetically closed structure of an ICME flux rope \citep{Demoulin09}. When modeled with polytropes, ICMEs show a non-adiabatic expansion with radial distance that suggests some local heating by turbulence \citep{Liu06} as in the solar wind generally. The radial evolution of the magnetic field strength, proton temperature, and density in ICMEs are such that low $\beta_i$ is maintained within ICMEs from the Sun to 1~au and beyond. This contrasts to the fast and slow winds, in which $\beta_i$ approaches unity well before reaching 1~au. 

Like other types of solar wind, ICMEs contain fluctuations at all measurable scales \citep[e.g.][]{Leamon98, Sorriso21, Marquez23} though at relatively small amplitudes \citep{Borovsky19}. The canonical power spectrum of fluctuations in the interplanetary magnetic field comprises distinctive power-law ranges \citep[e.g][]{Verscharen19}. These include a $k^{-1}$ wavenumber spectrum (`$1/f$ range' or `injection range') of non-interacting fluctuations at spatial scales exceeding the correlation length, thought to be an imprint from the wind's coronal sources. These large-scale fluctuations supply energy to a magnetohydrodynamic (MHD) cascade with a $k^{-3/2}$ or $k^{-5/3}$ spectrum between the correlation length and ion scales, the spectral indices in this `inertial range' being in agreement with various theories of Alfv\'enic MHD turbulence \citep{Schekochihin22}. Depending on the precise definition of compressibility used, typically 2-10\% of total fluctuation power is contained within compressive modes at MHD scales \citep{Chen16}. At ion scales, spectra with power-law indices significantly less than the $-7/3$ and $-8/3$ indices predicted for kinetic Alfv\'en wave turbulence may be observed; this range, which arises from a gradual transition between the MHD and kinetic regimes, is more consistently observed near the Sun than at 1~au \citep[][and references therein]{Bowen20}. At still smaller scales, ion-kinetic effects become fully developed and a $k^{-2.8}$ spectrum, most likely dominated by kinetic Alfv\'en wave turbulence, is present. 

In this Letter, we present spectral analysis of magnetic field fluctuations within ICMEs in terms of the phenomenology described above. For this analysis, ICMEs observed by the Solar Orbiter \citep[SolO;][]{Muller20} and Parker Solar Probe \citep[PSP;][]{Fox16} spacecraft have been selected, allowing the radial evolution of spectral properties in the inner heliosphere to be probed. A focus of the study has been to consider how certain well-known properties that distinguish ICME wind from other wind types -- namely, background fields with a flux rope geometry and low proton $\beta$ across a wide range of heliocentric distances -- may affect various spectral properties relating to MHD turbulence. Properties analyzed include fluctuation power and compressibility, spectral indices and break scales, and correlation lengths in the magnetic field. The rotation of the background flux rope field in ICME intervals often has a strong spectral signature at large scales, and care has been taken to remove this field when analyzing properties of the underlying large-scale fluctuations. More broadly, we consider whether ICMEs support turbulence that is different in nature to the turbulence present in non-ICME wind occupying a similar parameter regime (e.g. the near-Sun wind at low $\beta_i$), and, if there is some difference, whether the ICME environment could provide a new regime for understanding aspects of the turbulence.

\section{Data and Methods} \label{sec:data}

 The Helio4Cast ICMECAT database \citep{Mostl17,Mostl20} provides a regularly updated list of ICMEs observed by SolO and PSP. From the current list, a total of 28 ICME intervals with good magnetic field and plasma data coverage were identified and selected for analysis. Details of these ICMEs are listed in the Appendix. Only the magnetic driver intervals of the ICMEs have been analyzed and not other ICME substructures such as sheaths; all mentions of ICMEs in this work refer to the drivers only. The ICMEs were observed at heliocentric distances ranging from 0.25 to 0.95~au. All of the selected ICMEs had mean proton $\beta$ less than 0.4 and at least some coherent, flux-rope-like rotation of the large-scale magnetic field, signatures characteristic of magnetic clouds. The mean value of proton $\beta$ across all events was 0.16. Only the proton contribution to $\beta_i$ has been considered, and $\beta_i$ is hereafter equated with proton $\beta$. Magnetic field data from the SolO/MAG instrument \citep{Horbury20} and PSP/FIELDS instrument suite \citep{Bale16}, and ion moments from the SolO/SWA \citep{Owen20} and PSP/SWEAP-SPC instrument suites \citep{Kasper16}, have been used. The SolO/MAG data were at $\sim$0.127~s resolution and PSP/FIELDS data at 0.007 to 0.438~s resolution (typically 0.11~s), while data resolution from the plasma instruments varied between $\sim$1 and 28~s. Power spectral densities (PSDs) of magnetic field fluctuations in the ICME intervals have been determined using a multitapered fast-Fourier transform with bandwidth product NW~=~5/2. 

\section{Analysis} \label{sec:analysis}

\begin{figure}
\epsscale{1.19}
\plotone{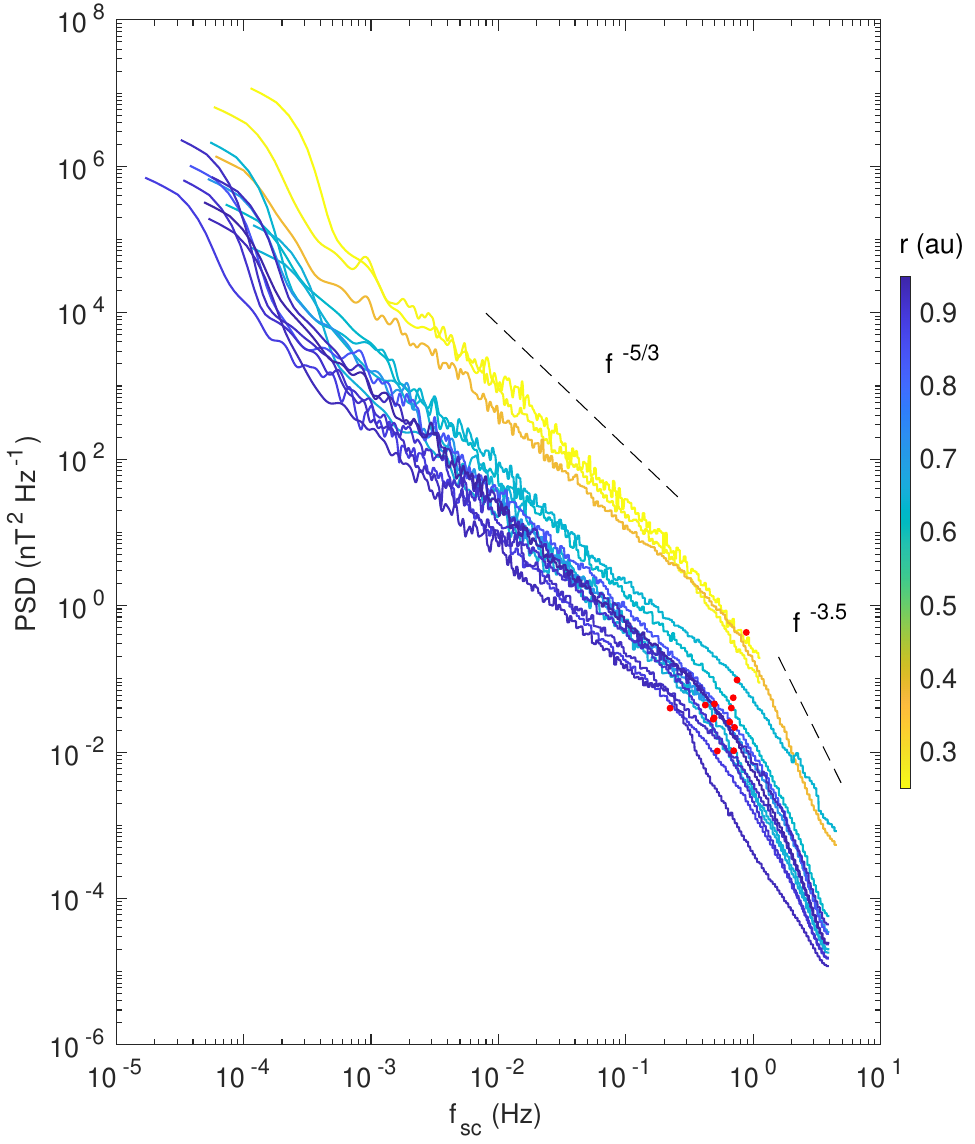}
\caption{Smoothed, trace PSD of magnetic field fluctuations for a selection of the ICMEs analyzed. Red points mark spectral break frequencies.}
\label{fig:spectra_distance}
\end{figure}

We begin this section by summarizing key features identified in the spectra, with the analysis described in detail in the following subsections. Figure~\ref{fig:spectra_distance} shows the smoothed, mean-subtracted, trace PSD (equivalent to the total power in fluctuations of the magnetic field vector) as a function of spacecraft-frame frequency, $f_{sc}$, for a selection of the ICMEs analyzed. The spectra are color-coded according to heliocentric distance, $r$. There is a general trend toward lower fluctuation power with increasing $r$ at all frequencies. A similar trend can be seen in an analogous figure with solar wind intervals by \citet{Chen20}, but with PSD lower in the ICMEs at most scales (e.g. by a factor of $\sim$5 at $f_{sc}=10^{-2}$~Hz) for a given $r$. The spectral indices are typically close to $-5/3$ at $f_{sc} \lesssim 10^{-1}$~Hz in the inertial range, with steepening of the spectral slopes characteristic of a transition toward the kinetic range evident at higher frequencies (Section~\ref{subsec:break_scales}). At $f_{sc} \lesssim 10^{-3}$~Hz, slopes are highly variable and in some cases significantly steeper than $-5/3$ due to power in the flux rope fields (Sections~\ref{subsec:break_scales} and \ref{subsec:icme_correlation}); only irregular signatures of an $1/f$ range are present at low frequencies. The spectral breaks at the high-frequency end of the inertial range, marked with the red points in Figure~\ref{fig:spectra_distance}, show an approximately linear tendency toward lower frequencies with $r$ (Section~\ref{subsec:evolution}).

\hfill

\subsection{Spectral Breaks and Slopes} \label{subsec:break_scales}

\begin{figure*}
\epsscale{1.18}
\plotone{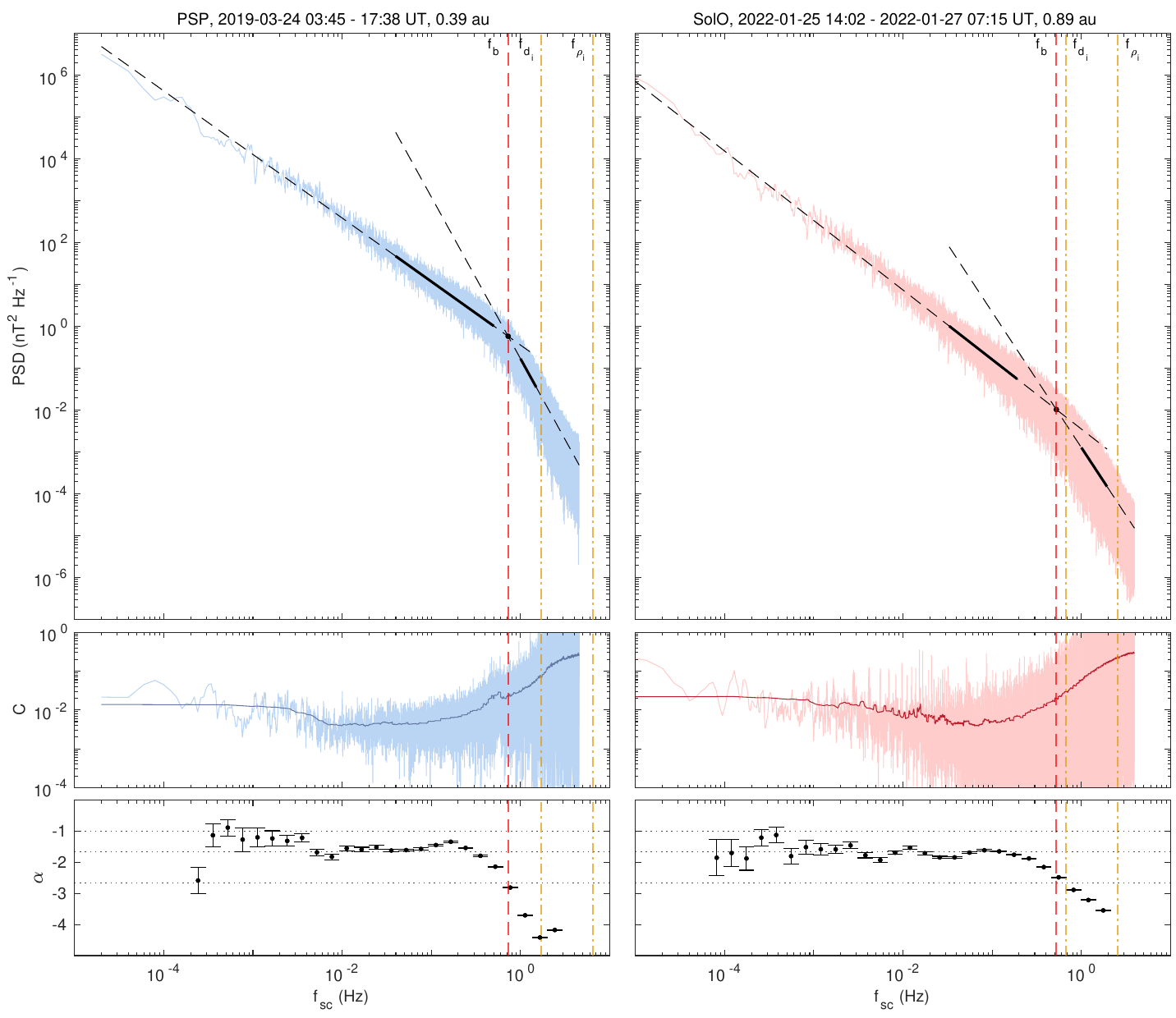}
\caption{Example spectra from two ICMEs. From top to bottom, the panels show the trace PSD with power-law fits in the inertial and transition ranges, magnetic compressibility spectrum, and spectral index of the trace PSD. Vertical lines mark spacecraft-frame frequencies associated with the spectral break point, $f_b$, the $k\rho_i=1$ scale, $f_{\rho_i}$, and the $kd_i=1$ scale, $f_{d_i}$. Horizontal lines in the bottom panels mark characteristic $\alpha$ values ($-1$, $-5/3$, and $-2.8$).}
\label{fig:spectra_examples} 
\end{figure*}

The top panels in Figure~\ref{fig:spectra_examples} show the trace PSD for two of the ICMEs, one observed by PSP at 0.39~au and the other by SolO at 0.89~au. Solid black lines represent linear fits to log(PSD) as a function of log($f_{sc}$) in the inertial and transition ranges. These fits have been extrapolated until they intersect, with the intersection point giving the spectral break frequency, $f_b$. The fits were applied to linear regions of the logarithmically scaled spectra such that $f_b$ approximately coincides with the midpoint of the frequency range where the spectral slope, $\alpha$, steepens in the transition range. The bottom panels in Figure~\ref{fig:spectra_examples} show $\alpha$ calculated using a sliding window, with $\alpha$ values given by the gradients of linear fits across the sliding-window range and uncertainties determined from the linear fit quality. For both ICMEs, it can be seen that $\alpha$ mostly fluctuates between $-1$ and $-2$ at $f_{sc} \leq 10^{-2}$~Hz, is approximately $-5/3$ at $10^{-2}$~Hz~$\leq f_{sc} \leq 10^{-1}$~Hz, and steepens to values below $-3$ at higher frequencies. The spectra of magnetic compressibility, $C$, that are shown in the middle panels are discussed in Section~\ref{subsec:compressibility}.

Spacecraft-frame frequencies associated with the proton gyroscale, $\rho_i$, and the proton inertial length, $d_i$, are marked in the figure. Applying Taylor's hypothesis, which gives wavenumber $k=2\pi f_{sc}/\langle v \rangle$ in the plasma frame, it may be shown that 
\begin{equation}
    k\rho_i = \frac{2\pi \sqrt{2 k_B \langle T_i \rangle m_i}}{e\langle v \rangle \langle B \rangle} f_{sc}
\end{equation}
and
\begin{equation}
    kd_i = \frac{2\pi}{e\langle v \rangle} \sqrt{\frac{m_i}{\mu_0 \langle n \rangle}} f_{sc},
    \label{eqn:kdi}
\end{equation}
where $T_i$, $m_i$, $e$, and $v$ are the proton temperature, mass, charge and speed, respectively, and angle brackets denote interval time averages. Thus spacecraft-frame frequencies $f_{\rho_i}$ and $f_{d_i}$ in Figure~\ref{fig:spectra_examples} correspond to the scales at which $k\rho_i=1$ and $kd_i=1$, respectively\footnote{Some authors compare $f_b$ to spacecraft-frame frequencies $f_l$ that relate to characteristic length scales $l$ via $kl=2\pi$ \citep[e.g.][]{Leamon00}, which would give corresponding frequencies that are a factor of $2\pi$ larger than $f_{\rho_i}$ and $f_{d_i}$ as defined here.}. It can be seen that $f_b$ is closer to $f_{d_i}$ than to $f_{\rho_i}$ in both of the examples shown. 

\begin{figure}
\epsscale{1.19}
\plotone{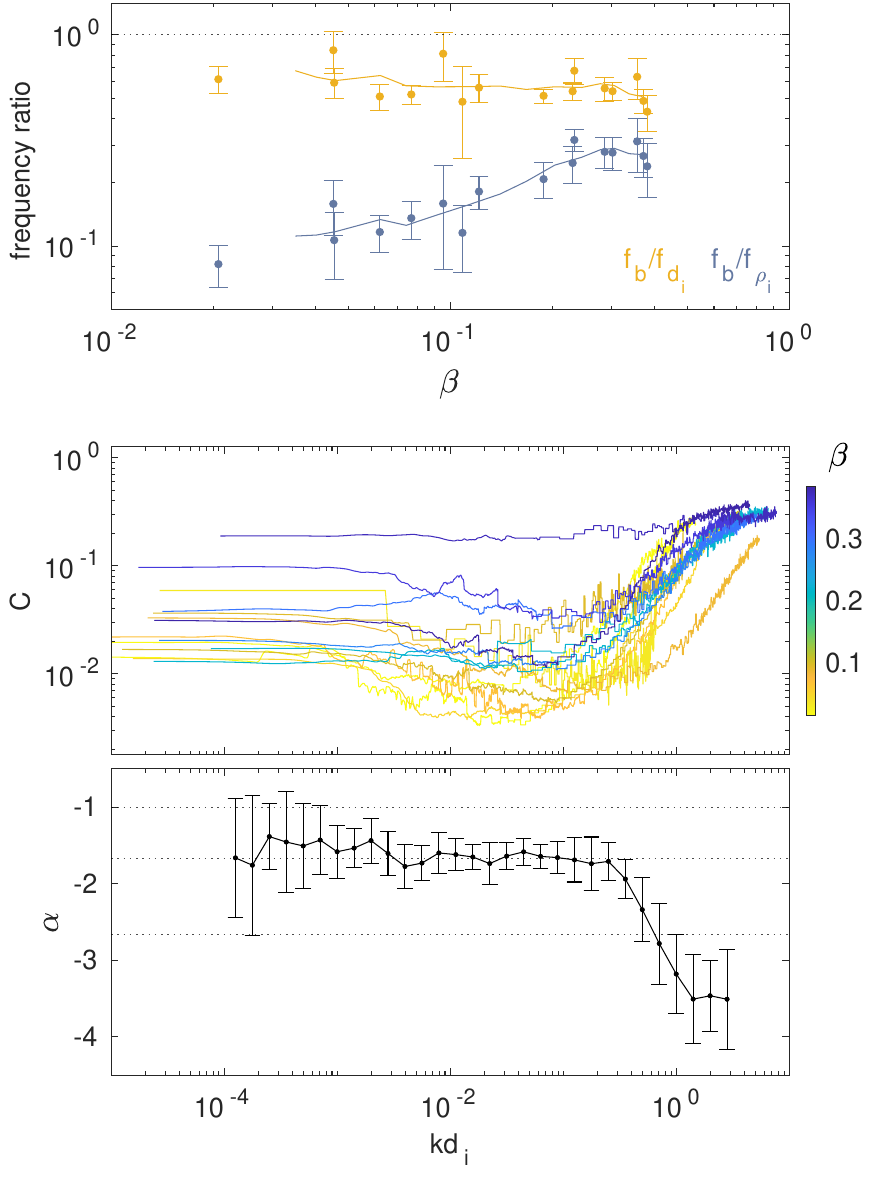}
\caption{Ratios of $f_b$ to $f_{d_i}$ and $f_{\rho_i}$ as a function of $\beta_i$ with five-point running average lines (top panel), smoothed compressibility spectra (middle panel), and mean spectral index (bottom panel) with characteristic values ($-1$, $-3/2$, $-5/3$, and $-2.8$) marked by horizontal lines.}
\label{fig:break_ratio_index} 
\end{figure}

Of the 28 ICME intervals analyzed, 16 had spectra without any significant non-power-law features (e.g. spikes, humps, or noise-floor flattening) near the spectral break. For these 16 spectra, $f_b$ could be accurately determined. The top panel of Figure~\ref{fig:break_ratio_index} shows the ratio of $f_b$ to $f_{\rho_i}$ and $f_{d_i}$ for the 16 events, as a function of the mean $\beta_i$ value in each ICME. Across the range of relatively low $\beta_i$ examined, $f_b/f_{d_i}$ was nearer unity than $f_b/f_{\rho_i}$, but with $f_b/f_{\rho_i}$ approaching unity as $\beta_i$ increased. This trend is in agreement with the findings of \citet{Chen14}, who determined that $d_i$ is systematically closer to the observed break scale at low $\beta_i$ and $\rho_i$ is closer at high $\beta_i$.

In order to make comparisons in the same reference frame and in terms of more physically meaningful units \citep[e.g.][]{Wicks10a,Sioulas23a}, all of the spectra are henceforth considered as a function of $kd_i$ rather than $f_{sc}$, with conversion from $f_{sc}$ given by Equation~\ref{eqn:kdi}. The bottom panel of Figure~\ref{fig:break_ratio_index} shows the sliding-window profile of $\alpha$ across the inertial and ion-kinetic ranges (\mbox{$10^{-4} < kd_i \leq 3$}), averaged across all events. Standard deviations are indicated by error bars. It can be seen that $\alpha$ is near $-5/3$ at $kd_i \lesssim 0.2$ in the inertial range and steepens to approximately $-3.5$ at $kd_i > 1$ in the kinetic range, but with high variability between events in the latter case. Variability is similarly large at $kd_i \lesssim 3 \times 10^{-3}$, where there is some indication of $\alpha$ being closer to $-3/2$ than $-5/3$. However, single fits at $kd_i \leq 10^{-3}$ that include points below the $kd_i \simeq 10^{-4}$ cutoff in Figure~\ref{fig:break_ratio_index} give a mean $\alpha \simeq -5/3$. The nature of the spectrum at low $kd_i$ is further considered in Section ~\ref{subsec:icme_correlation}. We note also that none of the trends identified in Figure~\ref{fig:break_ratio_index} showed a significant dependence on $r$. Radial evolution is considered explicitly in Section~\ref{subsec:evolution}.

\subsection{Compressibility Spectra} \label{subsec:compressibility}

The middle panels of Figure~\ref{fig:spectra_examples} show the magnetic compressibility, $C = (\delta|\textbf{\textit{B}}|/|\delta \textbf{\textit{B}}|)^2$, here obtained by dividing the PSD of fluctuations in the magnetic field magnitude by the trace PSD \citep[e.g.][]{Telloni21,Zhao21}. Compressibility gives the fraction of total fluctuation power in non-Alfv\'enic compressions as measured by $\delta|\textbf{\textit{B}}|$. The smoothed spectra of $C$ given by the darker overlaying lines show the same trend in both ICMEs: a broad minimum centered in the inertial range and rising values through the transition range, with the rise beginning at $f_{sc} < f_b$. A gradual plateauing of the rise can be seen between $f_{d_i}$ and $f_{\rho_i}$ with $C$ reaching $\sim$1/3 at the highest frequencies, this value being consistent with variance isotropy saturation \citep[e.g.][]{Matteini20}. The $C$ values in the inertial range are similar to those typically seen in non-ICME wind in the inner heliosphere \citep{Chen20}, but with $\delta \textbf{\textit{B}}$ and $\delta|\textbf{\textit{B}}|$ power both at lower levels in the ICMEs.

Smoothed compressibility spectra for 15 events are shown in the middle panel of Figure~\ref{fig:break_ratio_index}, color-coded to $\beta_i$. A moderate correlation between $C$ and $\beta_i$ that is qualitatively consistent with an expected relationship (discussed in Section~\ref{sec:discussion}) can be seen in Figure~\ref{fig:break_ratio_index} at inertial scales. The spectra generally take sigmoid forms for $kd_i \gtrsim 0.1$, converging to $C\sim 1/3$ at $kd_i \gtrsim 3$.

\subsection{Correlation Lengths and Flux Rope Fields} \label{subsec:icme_correlation}

\begin{figure*}
\epsscale{1.17}
\plotone{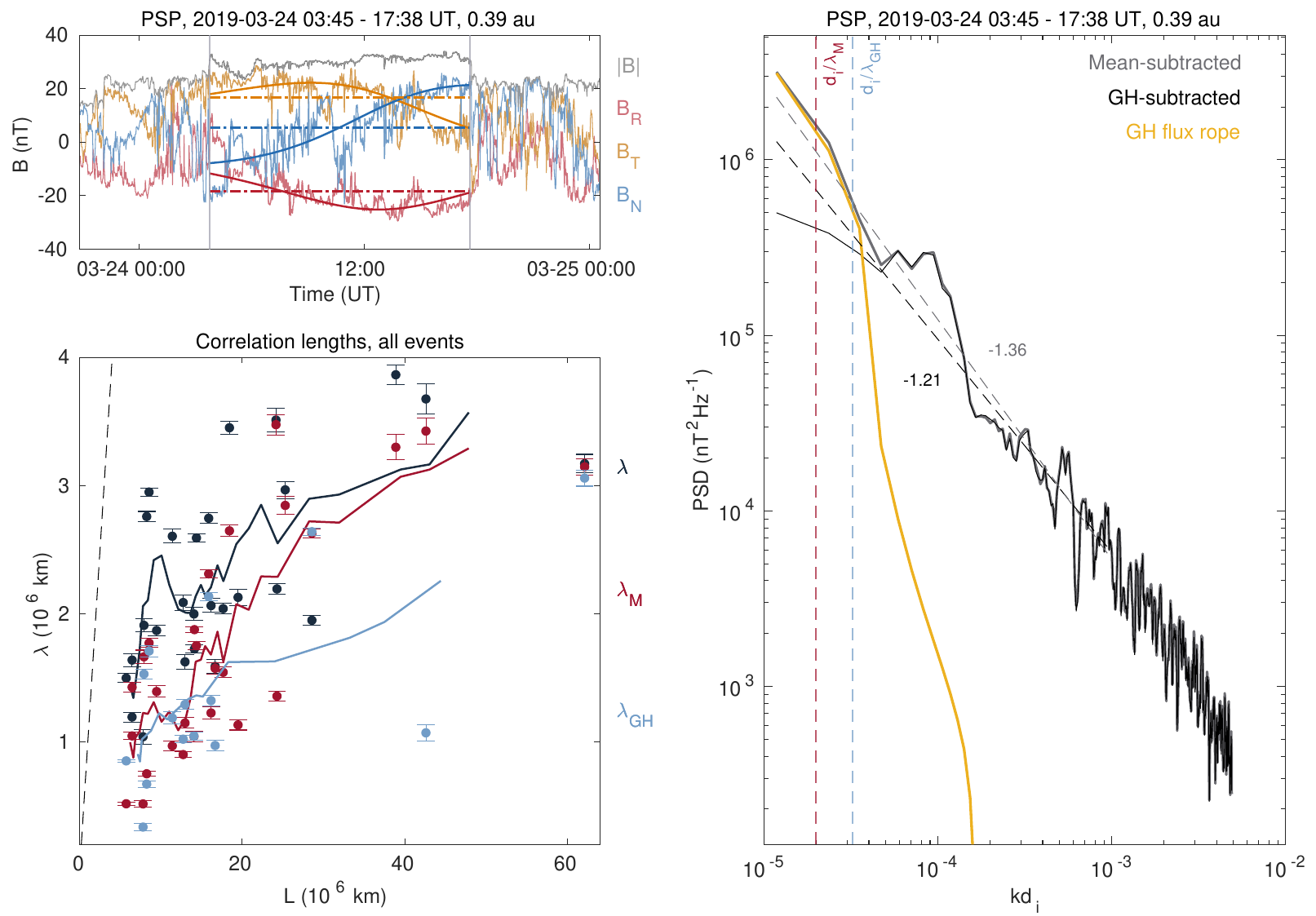}
\caption{Low-frequency analysis. The top-left panel show the magnetic field timeseries for an example ICME, overlaid with dash-dotted lines indicating the component mean values and solid lines showing a GH fit. Vertical lines mark the ICME boundaries. The right panel shows trace PSDs of the mean-subtracted, GH-subtracted, and modeled GH flux rope field. The bottom-left panel shows $\lambda$, $\lambda_M$, and $\lambda_{GH}$ versus ICME radial width, $L$, with five-point running averages shown by the smooth lines and $y=x$ by the dashed line.}
\label{fig:correlation_lengths} 
\end{figure*}

We now turn to the low-frequency limit of the inertial range, which is often found at scales near the correlation length \citep[e.g.][]{Matthaeus82, Burlaga84}. Magnetic field correlation lengths, $\lambda$, have been determined for 27 of the ICMEs using data at 1~min resolution. The ICME with the shortest duration (197~min) has been excluded from this analysis. For each ICME, the autocorrelation of a 224-min interval starting from the leading edge has been calculated, with lags, $\tau$, ranging from 0 to 112~min. The interval length has been chosen to equal two-thirds the duration of the shortest-duration ICME analyzed. An autocorrelation curve for each field component $j=\{R, T, N\}$ has been obtained and replotted as a function of spatial scale $l=\langle v \rangle \tau$, where $\langle v \rangle$ is the mean proton speed in the ICME. The curves have then been fitted with exponentials of the form $e^{-l/\lambda_j}$, with the average of the three $\lambda_j$ values, $\lambda$, taken as the overall value for the ICME interval. A similar technique has been used by \citet{Wicks10b}.

The correlation length determined in this way is likely sensitive to the spatial variation of the ICME's flux rope structure. To reduce this sensitivity, and thus to estimate the correlation length of the fluctuations rather than the background structure, correlation lengths have also been calculated with subtraction of the background field from the timeseries before performing the autocorrelation. The background field has been estimated with (i) the mean of each field component in the interval and (ii) with a Gold-Hoyle (GH) fit to the ICME flux rope \citep{Farrugia99}, giving correlation lengths $\lambda_M$ and $\lambda_{GH}$, respectively. The axial, poloidal, and radial components of the GH flux rope in cylindrical coordinates are given by
\begin{subequations}
\begin{equation}
B_z(\rho) = B_0 / (1 + \Gamma^2 \rho^2)
\end{equation}
\begin{equation}
B_{\phi}(\rho) = \Gamma\rho B_0 / (1 + \Gamma^2 \rho^2)
\end{equation}
\begin{equation}
B_{\rho}(\rho) = 0
\end{equation}
\end{subequations}
respectively, where fit-obtained constants $B_0$ and $\Gamma$ are the field magnitude at the axis and the field-line twist per unit length, respectively, and $\rho$ is the radial distance from the axis. Of the 27 ICMEs analyzed, 15 could be well fitted with the GH rope model. Key parameters of the 15 fits are listed in the Appendix. Further details of the GH fitting procedure are described by \citet{Kilpua19} and \citet{Good19}.

The top-left panel of Figure~\ref{fig:correlation_lengths} shows the magnetic field timeseries of an example ICME in RTN coordinates, with horizontal lines corresponding to the component mean values and smooth lines to the GH fit profile. The rotation of the magnetic field at the scale of the ICME duration is well captured by the GH fit, and is a better approximation of the ICME's structure than the component mean values. The bottom left panel in Figure~\ref{fig:correlation_lengths} shows the trace PSD for the same ICME with subtraction of the mean field (i.e. PSD calculated in the usual way) and with subtraction of the GH fit. While the mean-subtracted and GH-subtracted spectra overlap at $kd_i \gtrsim 10^{-4}$, it can be seen that power at the lowest frequencies (equivalent to the largest spatial scales) is reduced in the GH-subtracted spectrum, with a corresponding reduction in the spectral index from $-1.36$ to $-1.21$ at $kd_i \leq 10^{-3}$. Across all ICMEs fitted with the GH model, the average spectral index reduced from $-1.65$ in the mean-subtracted spectra to $-1.53$ in the GH-subtracted spectra at $kd_i \leq 10^{-3}$. Also displayed in Figure~\ref{fig:correlation_lengths} is the PSD of the GH fit profile (gold line, right panel), which shows the power contained within the low-frequency, large-amplitude rotation of the flux rope field. The significantly enhanced power and steep slopes seen at low frequencies in Figures \ref {fig:spectra_distance} and \ref{fig:spectra_examples} are due to such fields.

The bottom-left panel in Figure~\ref{fig:correlation_lengths} shows the three ICME correlation lengths versus ICME radial width, $L=\langle v \rangle T$, where $T$ is the ICME duration. The mean values of $\lambda$, $\lambda_M$ and $\lambda_{GH}$ are 2.3, 1.8 and $1.4 \times 10^{6}$~km, respectively, with $\lambda$ also being larger than both $\lambda_M$ and $\lambda_{GH}$ in all ICMEs individually. All three ICME correlation lengths tend to increase with $L$, and in all cases are significantly less than $L$. Five-point running average lines indicate some weak levelling-off in the increase in correlation lengths at large $L$, consistent with a nonlinear trend.

\subsection{Radial Evolution} \label{subsec:evolution}

\begin{figure}
\epsscale{1.15}
\plotone{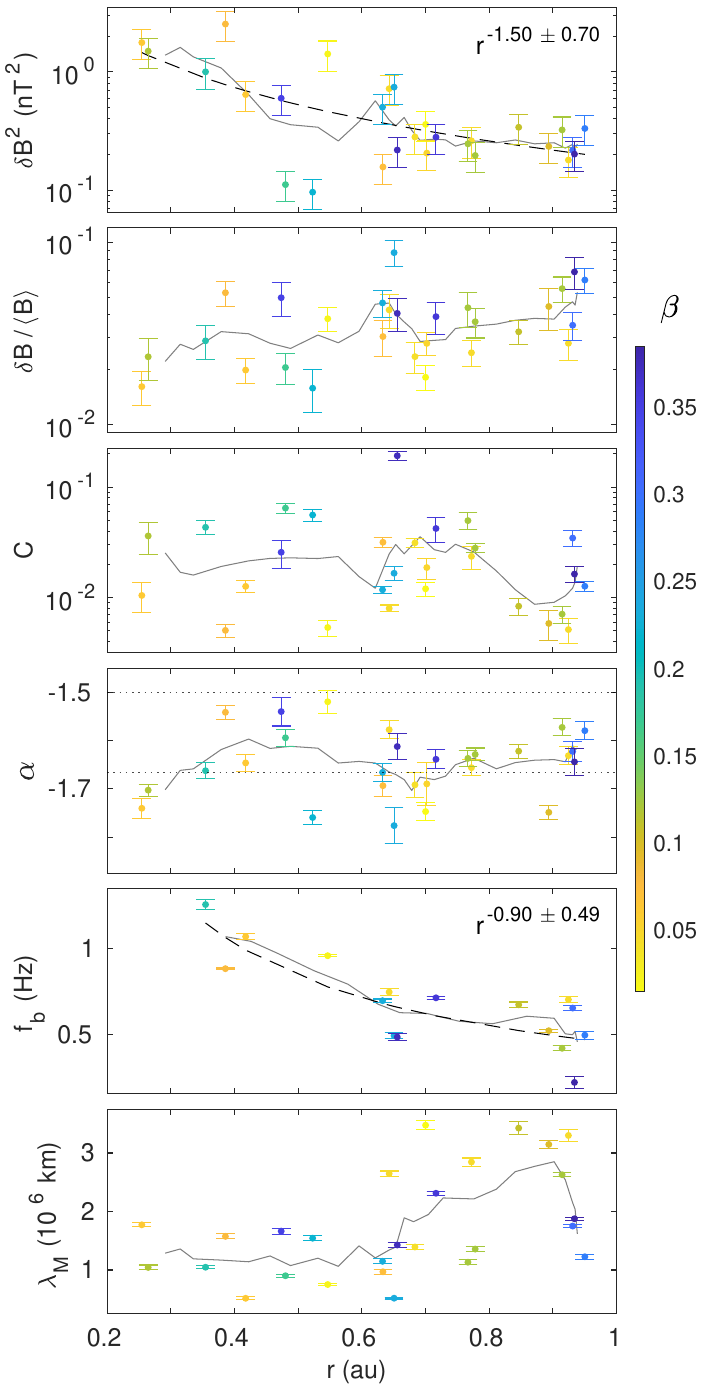}
\caption{Radial evolution of magnetic field parameters within the ICME intervals, color-coded to $\beta_i$. From top to bottom, the panels show the integrated fluctuation power, fluctuation amplitude normalized to the mean field strength, compressibility, inertial-range spectral index, spectral break frequency, and mean-subtracted correlation length. Gray lines show five-point moving averages and black dashed lines show power-law fits as a function of $r$.}
\label{fig:radial_evolution} 
\end{figure}

Figure~\ref{fig:radial_evolution} displays a range of parameters in the ICME intervals plotted versus $r$, with points color-coded to $\beta_i$. Values in the top four panels are averages in the inertial range at $10^{-2} \leq kd_i \leq 10^{-1}$. The gray lines show five-point moving averages for each parameter. Furthermore, correlations of the parameters with $r$ and $\beta_i$ have been calculated using the weighted Kendall coefficient, $\tau_{wK}$, as implemented by \citet{Lin18}. Similarly to other non-parametric correlation coefficients, perfect correlation is indicated by $\tau_{wK}=1$, perfect anticorrelation by $\tau_{wK}=-1$, and progressively weaker correlations or anticorrelations as $\tau_{wK}$ approaches zero. A weighted coefficient has been used because the parameters are generally heteroscedastic (i.e. have variances that change) with respect to $r$ and $\beta_i$.

Power-law fits as functions of $r$ are shown in Figure~\ref{fig:radial_evolution} for the two parameters that have the greatest $r$ correlation as measured by $|\tau_{wK}|$, namely $\delta B^2$ ($\tau_{wK}=-0.18$) and $f_b$ ($\tau_{wK}=-0.20$); the logarithm of these two parameters also show the greatest $r$ correlation ($\tau_{wK}=-0.24$ and $\tau_{wK}=-0.21$, respectively). The anticorrelation of $\delta B^2$ with $r$ and the $\delta B^2 \propto r^{-1.50}$ dependence reflect the falling PSD seen in Figure~\ref{fig:spectra_distance}. The $f_b\propto r^{-0.90}$ dependence is, within the fit uncertainty, comparable to the $r^{-1.09}$ dependence found by \citet{Bruno14} in fast wind and the $r^{-1.11}$ dependence found by \citet{Duan20} in slow wind. There is very broadly an increase in $\lambda_M$ with $r$, although the spread in values is large and the correlation with $r$ relatively weak ($\tau_{wK}=0.11$). The mean of $\lambda_M$ values at $r > 0.9$~au is in approximate agreement with the mean value of $2.33 \times 10^{6}$~km found by \citet{Ruiz14} in ICMEs at 1~au. The weakest correlations with $r$ are shown by $C$ and $\alpha$, both having $|\tau_{wK}|\lesssim0.05$. The correlation of $\beta_i$ with $r$ is negligible ($\tau_{wK}=0.01$), as expected for ICMEs.

The $\beta_i$ correlations measured by $|\tau_{wK}|$ are generally weak across all parameters, in some cases (e.g. $C$) possibly due to the narrow range of $\beta_i$ that has been examined. Of the parameters displayed in Figure~\ref{fig:radial_evolution}, the fluctuation amplitude normalized to the mean field strength across the interval, $\delta B / \langle B \rangle = \sqrt{\delta B^2} / \langle B \rangle$, has the greatest correlation with $\beta_i$ ($\tau_{wK}=0.14$, rising to $\tau_{wK}=0.18$ for the parameter logarithm). This correlation is consistent with a suppression of magnetic fluctuations at low $\beta_i$. As in Figure~\ref{fig:break_ratio_index}, the $\beta_i$ color map indicates a qualitative correlation between $C$ and $\beta_i$, although the quantified correlations are relatively low in this sample ($\tau_{wK}=0.10$ for the parameter logarithm).

\section{Discussion} \label{sec:discussion}

At large scales ($kd_i \leq 10^{-3}$), removal of the flux rope fields rather than the mean fields reduced the mean spectral slope to $-3/2$ from $-5/3$. Subtraction of the flux rope field is a form of detrending or frequency filtering that removes power in the spectrum associated with rotation of the background field, a signature of the flux rope transit over the spacecraft. Time-varying background fields in ICME intervals contrast with the approximately constant, Parker-spiral background in the solar wind more generally. It has been implicitly assumed that the flux rope field is a passive background that does not participate in the turbulent cascade, with the residual timeseries representing waves, turbulence, or structures not related to the global flux rope. The shallower slopes found at large scales with subtraction of the flux rope field may represent a distinct sub-region of the inertial range \citep[e.g.][]{Telloni22,Sioulas23b} or a roll-over toward a $1/f$ range. Sliding-window estimates of the spectral index in some cases do show $\alpha \simeq -1$ values at large scales with or without flux rope subtraction, but only sporadically. The properties of the fluctuations at these scales merit further investigation given their potential role in supplying energy to the turbulent cascade at smaller scales.

Correlation lengths of $\sim$4-hr intervals were longer in the total ICME fields than in the residual fields (mean or flux-rope-subtracted). Longer correlation lengths in the total field were likely due to the dominant influence of the background field rotation, with strongly autocorrelated variations at timescales of order $\sim$1 hr. ICMEs not displaying simple rotations in the background field may also have ordered global structures with relatively long correlation lengths. The correlation lengths in the background-subtracted fields may be related to the outer scale of turbulence, or, alternatively, to some typical mesoscale length of ICME substructure \citep{Lugaz18}. Taking the mean-subtracted correlation length, $\lambda_M$, as the turbulent outer scale and the high-frequency break scale calculated with Taylor's hypothesis, $l_b$, as the inner scale, then the span of the inertial range defined as $\lambda_M/l_b$ was approximately invariant with $r$ and $\beta_i$, at a constant value of $2.0 \pm 1.2 \times 10^{4}$ in the ICMEs analyzed. Correlation lengths increased with radial distance (as in the solar wind generally) and with the ICME radial width, $L$.

At $kd_i \gtrsim 10^{-3}$, spectral indices in the ICMEs were the same in the mean and flux-rope-subtracted fields. The presence at inertial scales of a $-5/3$ index across a range of heliocentric distances suggests turbulent states that are already well developed close to the Sun, that develop independently of the specific energy injection processes occurring at larger scales, and that may be related to the low cross helicity commonly found in ICMEs \citep{Good20,Good22,Soljento23}. Low cross helicity and an associated $-5/3$ index \citep{Podesta10} may in turn be general features of solar wind originating from closed-field regions in the corona \citep[e.g.][]{Borovsky19}. For example, solar wind near the heliospheric current sheet (HCS) has also been observed to have a $-5/3$ index close to the Sun \citep{Chen21}, in contrast to the more typical $-3/2$ value seen further from the HCS at the same distances \citep[e.g.][]{Chen20}; the radial evolution from a $-3/2$ to $-5/3$ index seen in the solar wind away from the HCS is likely caused by an evolution in cross helicity from highly imbalanced to balanced values \citep{Sioulas23a,McIntyre23}. Like most ICMEs, the HCS forms at the closed-field streamer belt, and ICMEs often locally replace the HCS in situ \citep{Crooker98}: thus some similarity between inertial-range spectral properties of near-HCS and ICME plasma might be expected. Steepening of the spectral index to values significantly below $-2.8$ in the transition range was observed consistently at all radial distances, unlike in the solar wind more generally. This behavior may be related to the ICMEs' low $\beta_i$, a dependency recently highlighted by \citet{Matteini20}.

The approximate invariance of magnetic compressibility with $r$ is notable. It has been suggested by \citet{Verscharen17} that the small amount of compressive fluctuation power in the solar wind is primarily due to the MHD slow mode rather than the kinetic slow mode. By assuming that the majority incompressible power is due to the MHD Alfv\'en mode, \citet{Chen20} find that the magnetic compressibility is given by $C = (\epsilon^2 \beta_i \gamma \sin^4\theta_{kB})/2$, where $\epsilon$ is the Alfv\'en-to-slow-mode amplitude ratio, $\gamma$ is the adiabatic index, and $\theta_{kB}$ is the slow-mode propagation angle relative to the mean field. Making the further assumption that $\gamma$, $\theta_{kB}\sim 90^{\circ}$, $C$, and $\beta_i$ are all approximately invariant with $r$ leads to the conclusion that $\epsilon^2$ is also invariant, such that ICMEs are a plasma environment in which the slow mode is frozen relative to the Alfv\'en mode with heliocentric distance. This contrasts with the rising slow-mode fraction found by \citet{Chen20} in the solar wind more generally. Furthermore, while the strong expansion and closed magnetic structuring described in Section~\ref{sec:intro} are likely the dominant influences, the particularly low power in compressive modes, which are a source of ion heating at inertial scales \citep[][and references therein]{Schekochihin19}, may also partly contribute to the low proton temperatures observed in ICMEs. A better understanding of turbulent heating in ICMEs could provide more stringent constraints for modeling their thermodynamic evolution with radial distance, and thus lead to more accurate modeling of ICME expansion and propagation for space weather prediction purposes.

\section{Conclusion} \label{sec:conclusion}

We have analyzed power spectra of magnetic field fluctuations within 28 ICMEs observed throughout the inner heliosphere by the PSP and SolO spacecraft. Low $\beta_i$ across a wide range of heliocentric distances and a global magnetic flux rope structure make ICMEs a distinctive space plasma environment. At large spatial scales comparable to the correlation length ($kd_i\lesssim 10^{-4}$), a significant fraction of power was contained within the background flux rope fields of the ICMEs. Subtraction of these fields from the timeseries revealed shorter correlation lengths and shallower spectral slopes that on average reduced from $-5/3$ to $-3/2$ at $kd_i\leq 10^{-3}$, with some sporadic signatures of a $-1$ slope also at these scales. The mean value of the spectral slope deep in the inertial range was $-5/3$ and did not show the $-3/2$ to $-5/3$ radial evolution observed in non-ICME wind away from the HCS. The inertial range terminated at scales closer to the proton inertial length than the proton gyroscale, consistent with previous analysis of low-$\beta_i$ plasma. Steepening of the spectral slope to values below $-3$ in the transition range was observed at all radial distances. Magnetic compressibility in the inertial range held similar values to the non-ICME wind in the inner heliosphere but did not grow with radial distance to 1~au, likely mirroring very weak radial variations in $\beta_i$ and the Alfv\'en-to-slow-mode fraction. Scaling of the Alfv\'enic turbulence in the inertial range appears to be independent of the global flux rope structure and low $\beta_i$ that characterize ICMEs, suggestive of its universal, system-independent nature.

\begin{acknowledgments}
This work was funded by an Academy of Finland Research Fellowship (grants 338486 and 346612; \mbox{INERTUM}). C.H.K.C. is supported by UKRI Future Leaders Fellowship MR/W007657/1 and STFC Consolidated Grants ST/T00018X/1 and ST/X000974/1. C.M. is funded by the European Union (ERC, HELIO4CAST, 101042188). E.K.J.K. acknowledges support from Academy of Finland Centre of Excellence \mbox{FORESAIL} (grant 336807) and from the European Research Council under the European Union’s Horizon 2020 research and innovation programme, grant 724391 (SolMAG). Views and opinions expressed are those of the authors only and do not necessarily reflect those of the European Union or the European Research Council Executive Agency. Neither the European Union nor the granting authority can be held responsible for them. We thank the Solar Orbiter and Parker Solar Probe instrument teams for providing the data used in this work, and also the reviewers for their constructive comments on the manuscript. Open access to this work has been funded by Helsinki University Library.
\end{acknowledgments}

\clearpage

\appendix
\label{appendix:event_table}

The start and end times of the ICMEs analyzed in this study are listed in Table~\ref{table:events}, and are taken from the Helio4Cast ICMECAT database (https://helioforecast.space/icmecat). For those ICMEs that have been fitted with the Gold-Hoyle flux rope model, key parameters of the fits are also listed in the table. The fit parameters include: impact parameter of the spacecraft with the flux rope, $p$ (`0' indicating an axis encounter, `1' an edge encounter); the angle between the rope axis and the R--T plane, $\theta_0$; the angle between the projection of the rope axis onto the R--T plane and the R direction, $\phi_0$; the magnetic field strength at the rope axis, $B_0$; and the field-line twist per unit length, $\Gamma$.

\begin{table}[!h] \caption{List of ICMEs Analyzed}
\centering
\begin{tabular}{c c c c c c c c c c}
\hline\hline
Event number & Start time (UT) & End time (UT) & Spacecraft & $r$ (au) & $p$ & $\theta_0$ ($^{o}$) & $\phi_0$ ($^{o}$) & $B_0$ (nT) & $\Gamma$ (au$^{-1}$)\\ [0.5ex]
\hline

1&    2018-10-30 20:25   &	2018-10-31 08:19   &	PSP   &	0.26  	&	0.17	&	-11	&	153	&	56.0	&	-5.2   \\
2&    2018-11-11 23:51   &	2018-11-12 05:59   &	PSP   &	0.25  	&	0.21	&	18	&	79	&	100.6	&	7.3 \\
3&    2019-03-15 12:11   &	2019-03-15 17:49   &	PSP   &	0.55  	&	0.49	&	10	&	106	&	36.0	&	3.6 \\
4&    2019-03-24 03:45   &	2019-03-24 17:38   &	PSP   &	0.39  	&	0.58	&	31	&	119	&	36.7	&	3.0 \\
5&    2020-02-11 05:08   &	2020-02-11 11:35   &	PSP   &	0.42  	&	0.02	&	10	&	89	&	43.6	&	4.6  \\
6&    2020-05-28 08:50   &	2020-05-28 14:59   &	PSP   &	0.35    &   --      &   --  &   --  &   --      &   -- \\
7&    2020-06-23 07:07   &	2020-06-23 16:52   &	PSP   &	0.48  	&	0.70	&	39	&	310	&	20.1	&	-2.7    \\
8&    2020-06-25 15:59   &	2020-06-26 08:15   &	PSP   &	0.52  &   --      &   --  &   --  &   --      &   -- \\
9&    2020-09-12 13:33   &	2020-09-12 19:35   &	PSP   &	0.47  	&	0.56	&	32	&	322	&	23.7	&	7.3 \\
10&    2020-10-28 13:24   &	2020-10-28 16:41   &	PSP   &	0.70  &   --      &   --  &   --  &   --      &   -- \\
11&    2021-02-11 16:17   &	2021-02-12 00:57   &	PSP   &	0.63  	&	0.26	&	15	&	291	&	15.4	&	2.6 \\
12&    2021-02-12 11:17   &	2021-02-12 23:59   &	PSP   &	0.64   &   --      &   --  &   --  &   --      &   -- \\
13&    2021-05-06 18:26   &	2021-05-07 15:07   &	SolO   &	0.91 	&	0.48	&	18	&	250	&	13.6	&	-3.1   \\
14&    2021-05-10 14:01   &	2021-05-11 11:39   &	SolO   &	0.92 &   --      &   --  &   --  &   --      &   -- \\
15&    2021-05-27 20:14   &	2021-05-28 10:27   &	SolO   &	0.95 	&	0.06	&	2	&	274	&	11.1	&	4.0 \\
16&    2021-05-30 14:18   &	2021-05-31 00:44   &	PSP   &	0.70  &   --      &   --  &   --  &   --      &   -- \\
17&    2021-06-10 16:52   &	2021-06-11 08:53   &	PSP   &	0.77  &   --      &   --  &   --  &   --      &   -- \\
18&    2021-06-12 09:20   &	2021-06-13 01:12   &	PSP   &	0.77  &   --      &   --  &   --  &   --      &   -- \\
19&    2021-06-21 12:22   &	2021-06-22 01:29   &	SolO   &	0.93  &   --      &   --  &   --  &   --      &   -- \\
20&    2021-06-22 22:28   &	2021-06-23 10:50   &	SolO   &	0.93 &   --      &   --  &   --  &   --      &   -- \\
21&    2021-06-24 08:10   &	2021-06-25 06:05   &	PSP   &	0.78  &   --      &   --  &   --  &   --      &   -- \\
22&    2021-07-11 10:25   &	2021-07-11 18:31   &	PSP   &	0.68  &   --      &   --  &   --  &   --      &   -- \\
23&    2021-08-25 01:06   &	2021-08-25 10:43   &	SolO   &	0.63 	&	0.04	&	5	&	272	&	16.2	&	-2.3    \\
24&    2021-10-04 07:01   &	2021-10-04 12:59   &	SolO   &	0.65 	&	0.18	&	-7	&	285	&	11.6	&	-6.6   \\
25&    2021-10-05 05:09   &	2021-10-05 11:00   &	SolO   &	0.66 &   --      &   --  &   --  &   --      &   -- \\
26&    2021-10-15 09:54   &	2021-10-15 23:11   &	SolO   &	0.72 	&	0.01	&	-28	&	271	&	14.7	&	-2.2  \\
27&    2021-11-04 01:25   &	2021-11-04 19:47   &	SolO   &	0.83 	&	0.31	&	64	&	311	&	20.6	&	-1.0   \\
28&    2022-01-25 14:02   &	2022-01-27 07:15   &	SolO   &	0.89	&	0.12	&	-5	&	261	&	12.5	&	-0.7 \\ [1ex]

\hline 
\end{tabular} \label{table:events}
\end{table}

\bibliography{references}{}
\bibliographystyle{aasjournal}

\end{document}